\documentclass[12pt,a4paper]{article}

\usepackage{ifthen} 
\newboolean{pdflatex}
\setboolean{pdflatex}{true} 

\newboolean{articletitles}
\setboolean{articletitles}{true} 

\newboolean{uprightparticles}
\setboolean{uprightparticles}{false} 

\def\paperauthors{P. Billoir, M. De Cian, P. A. G\"unther, S. Stemmle} 
\def\paperasciititle{A parametrized Kalman filter for fast track fitting at LHCb} 
\def\papertitle{A parametrized Kalman filter for fast track fitting at \lhcb} 
\def\paperkeywords{{High Energy Physics}, {LHCb}} 
\def\papercopyright{\the\year\ CERN for the benefit of the LHCb collaboration} 
\def\paperlicence{CC-BY-4.0 licence}
\def\paperlicenceurl{https://creativecommons.org/licenses/by/4.0/}

\usepackage[top=1in, bottom=1.25in, left=1in, right=1in]{geometry}

\usepackage{microtype}
\usepackage{lineno}  
\usepackage{xspace} 
\usepackage{caption} 

\usepackage{graphicx}  
\usepackage{color}
\usepackage{colortbl}
\graphicspath{{./figs/}} 
\DeclareGraphicsExtensions{.pdf,.PDF,png,.PNG}

\usepackage{amsmath} 
\usepackage{amssymb}
\usepackage{amsfonts}
\usepackage{upgreek} 

\newcommand*\patchAmsMathEnvironmentForLineno[1]{%
\expandafter\let\csname old#1\expandafter\endcsname\csname #1\endcsname
\expandafter\let\csname oldend#1\expandafter\endcsname\csname
end#1\endcsname
 \renewenvironment{#1}%
   {\linenomath\csname old#1\endcsname}%
   {\csname oldend#1\endcsname\endlinenomath}%
}
\newcommand*\patchBothAmsMathEnvironmentsForLineno[1]{%
  \patchAmsMathEnvironmentForLineno{#1}%
  \patchAmsMathEnvironmentForLineno{#1*}%
}
\AtBeginDocument{%
\patchBothAmsMathEnvironmentsForLineno{equation}%
\patchBothAmsMathEnvironmentsForLineno{align}%
\patchBothAmsMathEnvironmentsForLineno{flalign}%
\patchBothAmsMathEnvironmentsForLineno{alignat}%
\patchBothAmsMathEnvironmentsForLineno{gather}%
\patchBothAmsMathEnvironmentsForLineno{multline}%
\patchBothAmsMathEnvironmentsForLineno{eqnarray}%
}

\usepackage{hyperxmp}

\usepackage[pdftex,
            pdfauthor={\paperauthors},
            pdftitle={\paperasciititle},
            pdfkeywords={\paperkeywords},
            pdfcopyright={Copyright (C) \papercopyright},
            pdflicenseurl={\paperlicenceurl}]{hyperref}

\usepackage[colorinlistoftodos,textsize=scriptsize]{todonotes}

\usepackage[all]{hypcap} 

\usepackage{xspace} 
\usepackage{upgreek}

\def\lhcb   {\mbox{LHCb}\xspace}

\def\lhc    {\mbox{LHC}\xspace}

\def\velo   {VELO\xspace}

\def\MagUp {\mbox{\em Mag\kern -0.05em Up}\xspace}

\ifthenelse{\boolean{uprightparticles}}%
{

 \def\PDelta      {\ensuremath{\Delta}\xspace}                 
 \def\PXi         {\ensuremath{\Xi}\xspace}                 
 \def\PLambda     {\ensuremath{\Lambda}\xspace}                 
 \def\PSigma      {\ensuremath{\Sigma}\xspace}                 
 \def\POmega      {\ensuremath{\Omega}\xspace}                 
 \def\PUpsilon    {\ensuremath{\Upsilon}\xspace}

 \def\PB      {\ensuremath{\mathrm{B}}\xspace}                 
                  
 \def\PD      {\ensuremath{\mathrm{D}}\xspace}

 \def\PK      {\ensuremath{\mathrm{K}}\xspace}

 \def\Pb      {\ensuremath{\mathrm{b}}\xspace}                 
 \def\Pc      {\ensuremath{\mathrm{c}}\xspace}

 \def\Pi      {\ensuremath{\mathrm{i}}\xspace}

 \def\Ps      {\ensuremath{\mathrm{s}}\xspace}

 \def\thebaroffset{0.0em}
}
{

 \mathchardef\PDelta="7101
 \mathchardef\PXi="7104
 \mathchardef\PLambda="7103
 \mathchardef\PSigma="7106
 \mathchardef\POmega="710A
 \mathchardef\PUpsilon="7107
                  
 \def\PB      {\ensuremath{B}\xspace}                 
                  
 \def\PD      {\ensuremath{D}\xspace}

 \def\PK      {\ensuremath{K}\xspace}

 \def\Pb      {\ensuremath{b}\xspace}                 
 \def\Pc      {\ensuremath{c}\xspace}

 \def\Pi      {\ensuremath{i}\xspace}

 \def\Ps      {\ensuremath{s}\xspace}

 \def\thebaroffset{0.18em}
}
\newcommand{\offsetoverline}[2][\thebaroffset]{\kern #1\overline{\kern -#1 #2}}%

\makeatletter
\ifcase \@ptsize \relax
  \newcommand{\miniscule}{\@setfontsize\miniscule{4}{5}}
\or
  \newcommand{\miniscule}{\@setfontsize\miniscule{5}{6}}
\or
  \newcommand{\miniscule}{\@setfontsize\miniscule{5}{6}}
\fi
\makeatother

\DeclareRobustCommand{\optbar}[1]{\shortstack{{\miniscule (\rule[.5ex]{1.25em}{.18mm})}
  \\ [-.7ex] $#1$}}

\def\squark    {{\ensuremath{\Ps}}\xspace}

\def\cquark    {{\ensuremath{\Pc}}\xspace}

\def\bquark    {{\ensuremath{\Pb}}\xspace}

\def\kaon    {{\ensuremath{\PK}}\xspace}

\def\KorKbar {\kern \thebaroffset\optbar{\kern -\thebaroffset \PK}{}\xspace}

\def\Kp      {{\ensuremath{\kaon^+}}\xspace}
\def\Km      {{\ensuremath{\kaon^-}}\xspace}

\def\DorDbar {\kern \thebaroffset\optbar{\kern -\thebaroffset \PD}\xspace}

\def\B       {{\ensuremath{\PB}}\xspace}

\def\BorBbar {\kern \thebaroffset\optbar{\kern -\thebaroffset \PB}\xspace}

\def\Bd      {{\ensuremath{\B^0}}\xspace}

\def\BdorBdbar {\kern \thebaroffset\optbar{\kern -\thebaroffset \Bd}\xspace}

\def\Bs      {{\ensuremath{\B^0_\squark}}\xspace}

\def\BsorBsbar {\kern \thebaroffset\optbar{\kern -\thebaroffset \Bs}\xspace}

\def\Y#1S{\ensuremath{\PUpsilon{(#1S)}}\xspace}

\def\LorLbar     {\kern \thebaroffset\optbar{\kern -\thebaroffset \PLambda}\xspace}

\newcommand{\decay}[2]{\ensuremath{#1\!\to #2}\xspace} 

\def\to                 {\ensuremath{\rightarrow}\xspace}

\def\AT#1     {\ensuremath{A_{\mathrm{T}}^{#1}}\xspace}           

\def\C#1      {\ensuremath{\mathcal{C}_{#1}}\xspace}                       
\def\Cp#1     {\ensuremath{\mathcal{C}_{#1}^{'}}\xspace}                    
\def\Ceff#1   {\ensuremath{\mathcal{C}_{#1}^{\mathrm{(eff)}}}\xspace}        
\def\Cpeff#1  {\ensuremath{\mathcal{C}_{#1}^{'\mathrm{(eff)}}}\xspace}       
\def\Ope#1    {\ensuremath{\mathcal{O}_{#1}}\xspace}                       
\def\Opep#1   {\ensuremath{\mathcal{O}_{#1}^{'}}\xspace}                    

\newcommand{\aunit}[1]{\ensuremath{\text{\,#1}}}       

\newcommand{\tev}{\aunit{Te\kern -0.1em V}\xspace}
\newcommand{\gev}{\aunit{Ge\kern -0.1em V}\xspace}
\newcommand{\mev}{\aunit{Me\kern -0.1em V}\xspace}
\newcommand{\kev}{\aunit{ke\kern -0.1em V}\xspace}
\newcommand{\ev}{\aunit{e\kern -0.1em V}\xspace}
\newcommand{\mevc}{\ensuremath{\aunit{Me\kern -0.1em V\!/}c}\xspace}
\newcommand{\gevc}{\ensuremath{\aunit{Ge\kern -0.1em V\!/}c}\xspace}
\newcommand{\mevcc}{\ensuremath{\aunit{Me\kern -0.1em V\!/}c^2}\xspace}
\newcommand{\gevcc}{\ensuremath{\aunit{Ge\kern -0.1em V\!/}c^2}\xspace}

\def\mhz  {\ensuremath{\aunit{MHz}}\xspace}

\def\gsim{{~\raise.15em\hbox{$>$}\kern-.85em
          \lower.35em\hbox{$\sim$}~}\xspace}
\def\lsim{{~\raise.15em\hbox{$<$}\kern-.85em
          \lower.35em\hbox{$\sim$}~}\xspace}

\def\pt         {\ensuremath{p_{\mathrm{T}}}\xspace}

\def\ptot       {\ensuremath{p}\xspace}

\def\evtgen     {\mbox{\textsc{EvtGen}}\xspace}

\def\geant      {\mbox{\textsc{Geant4}}\xspace}

\def\photos     {\mbox{\textsc{Photos}}\xspace}

\def\pythia     {\mbox{\textsc{Pythia}}\xspace}

\def\tell1  {TELL1\xspace}
\def\ukl1   {UKL1\xspace}

\newcommand{\eg}{\mbox{\itshape e.g.}\xspace}
\newcommand{\ie}{\mbox{\itshape i.e.}\xspace}

\usepackage{cite} 
\usepackage{mciteplus}

\begin{document}

\renewcommand{\thefootnote}{\fnsymbol{footnote}}
\setcounter{footnote}{1}

\begin{titlepage}
\pagenumbering{roman}

\vspace*{-1.5cm}
\centerline{\large EUROPEAN ORGANIZATION FOR NUCLEAR RESEARCH (CERN)}
\vspace*{1.5cm}
\noindent
\begin{tabular*}{\linewidth}{lc@{\extracolsep{\fill}}r@{\extracolsep{0pt}}}
\ifthenelse{\boolean{pdflatex}}
{\vspace*{-1.5cm}\mbox{\!\!\!\includegraphics[width=.14\textwidth]{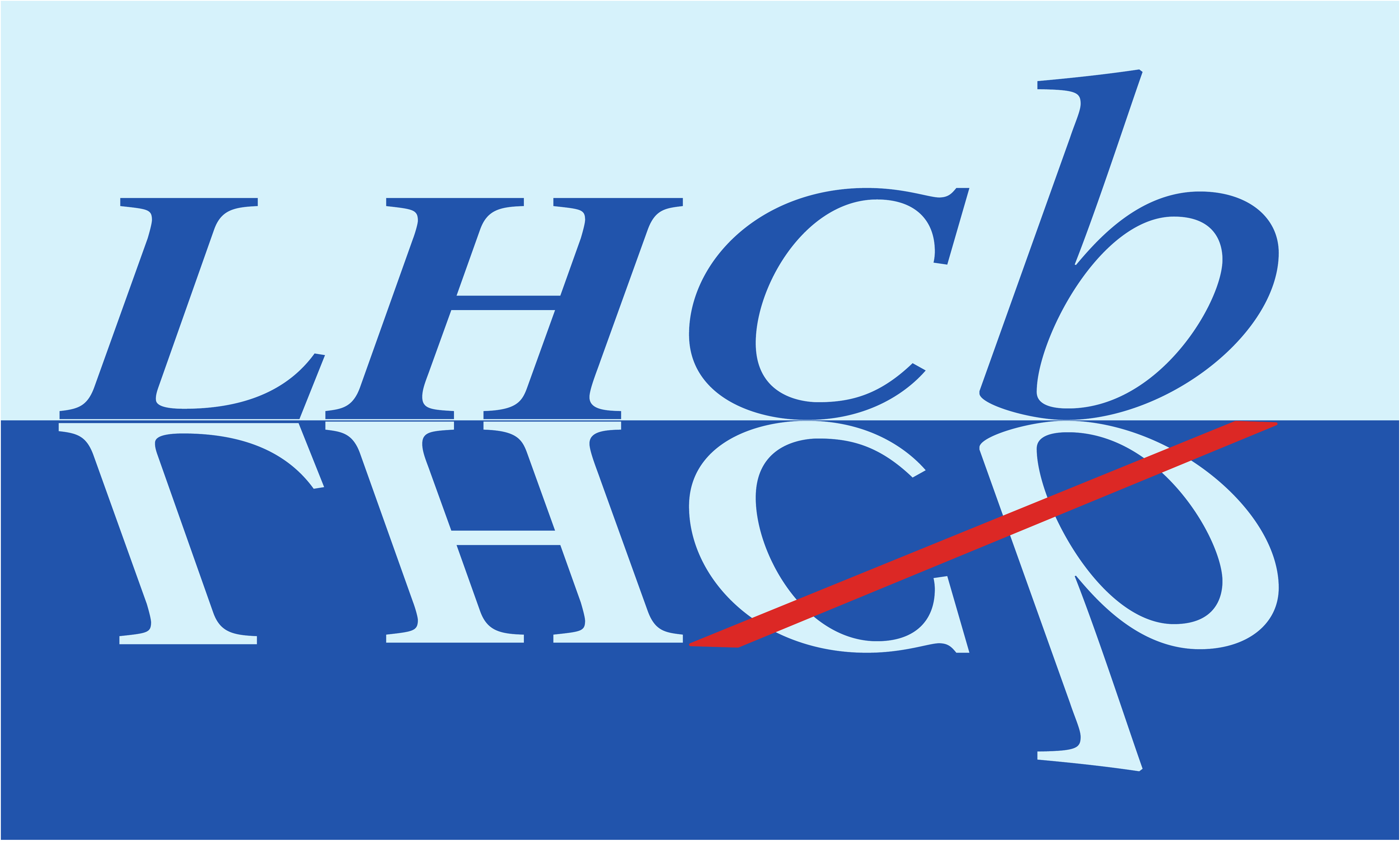}} & &}%
{\vspace*{-1.2cm}\mbox{\!\!\!\includegraphics[width=.12\textwidth]{figs/lhcb-logo.eps}} & &}%
\\
 & & LHCb-DP-2021-001 \\  
 & & May 17, 2021 \\
 & & \\
\end{tabular*}

\vspace*{4.0cm}

{\normalfont\bfseries\boldmath\huge
\begin{center}
  \papertitle 
\end{center}
}

\vspace*{2.0cm}

\begin{center}

P. Billoir$^1$, M. De Cian$^2$, P. A. G\"unther$^3$, S. Stemmle$^{3,\dagger}$
\bigskip\\
{\normalfont\itshape\footnotesize
$^1$LPNHE, Sorbonne Universit\'e, Paris Diderot Sorbonne Paris Cit\'e, CNRS/IN2P3, Paris, France\
$^2$Institute of Physics, Ecole Polytechnique F\'ed\'erale de Lausanne (EPFL), Lausanne, Switzerland\\
$^3$Physikalisches Institut, Ruprecht-Karls-Universit\"at Heidelberg, Heidelberg, Germany\\
$^{\dagger}$Author was at institute at time work was performed.
\\
}
\end{center}

\vspace{\fill}

\begin{abstract}
  \noindent
   We present an alternative implementation of the Kalman filter employed for track fitting within the LHCb experiment. It uses simple parametrizations for the extrapolation of particle trajectories
   in the field of the LHCb dipole magnet
   and for the effects of multiple scattering in the detector material. A speedup of more than a factor of four is achieved while maintaining the quality of the estimated track quantities. This Kalman filter implementation could be used in the purely software-based trigger of the LHCb upgrade.
  
\end{abstract}

\vspace*{2.0cm}
\begin{center}
Published in \href{https://www.sciencedirect.com/science/article/pii/S0010465521001387}{Computer Physics Communications \textbf{265}, 108026 (2021)}
\end{center}

\vspace{\fill}

{\footnotesize 
\centerline{\copyright~\papercopyright. \href{\paperlicenceurl}{\paperlicence}.}}
\vspace*{2mm}

\end{titlepage}


\newpage
\setcounter{page}{2}
\mbox{~}


\renewcommand{\thefootnote}{\arabic{footnote}}
\setcounter{footnote}{0}



\pagestyle{plain} 
\setcounter{page}{1}
\pagenumbering{arabic}


%

\section{Introduction}
\label{sec:Introduction}

The \lhcb experiment is a dedicated heavy flavour physics experiment at the \lhc focusing on the study of hadrons containing \bquark and \cquark quarks~\cite{LHCb-DP-2008-001}. Due to the high luminosity at the \lhc and the high proton-proton interaction cross section, a sophisticated trigger system is needed to reduce the rate of collisions saved for offline analysis. During Runs 1 and 2 of the \lhc, this trigger system consisted of a hardware stage, reducing the rate from $40\mhz$ to $1\mhz$, followed by a two-stage software trigger. In the latter, the full tracking system was read out and a partial (first stage) and full (second stage) event reconstruction were performed~\cite{LHCb-DP-2019-001}.
Both software stages included a fit of selected track candidates using a Kalman filter to extract their parameters and to reject fake tracks. In addition, the software trigger allowed an online calibration and alignment of the detector~\cite{BORGHI2017560}.

During Run 3 of the \lhc, \lhcb will be provided with a factor five higher luminosity compared to Run 2. In this scope, most of the subdetectors are currently being replaced or upgraded\cite{LHCb-TDR-012,LHCb-TDR-013,LHCb-TDR-014,LHCb-TDR-015} and a new trigger strategy has been developed \cite{LHCb-TDR-016}.
The hardware trigger will be removed and a two-stage, fully software-based trigger will process the full $30\mhz$\footnote{The nominal bunch-crossing frequency of the \lhc is 40\mhz, however empty and non-colliding bunches reduce this to a collision frequency of 30\mhz at \lhcb.} of bunch-crossing rate. In the first stage, tracks with a high transverse momentum (\pt) and primary vertices will be reconstructed. These objects are used to select events with displaced topologies typical for \bquark-hadron and \cquark-hadron decays, and to select high-\pt objects from decays of heavy vector bosons.
In the second stage, a full event reconstruction will be performed, without any requirement on the \pt and including particle identification. 
A large number of exclusive and several universal event selections based on the decay topology will be applied.

In \lhcb, track reconstruction is split into a \textit{pattern recognition} and a \textit{Kalman filtering}\cite{Kalman1960, FRUHWIRTH1987444} stage. 
During pattern recognition, sets in each subdetector are constructed from signals that potentially result from the passage of a single charged particle. Simple parametrizations are used throughout this procedure as it is only concerned with finding the right sets of signals and not to provide the best estimate of the track parameters. During the filtering stage, an estimate for the track parameters is calculated, and fake tracks are rejected. Given that the output of the filtering stage is used for physics selections the best possible precision needs to be achieved, hence an (extended) Kalman filter is used for track fitting.
Ideally, Kalman filtering of the track candidates is already performed during the first trigger stage. However, the Kalman filter which was used during Run 1 and 2 in \lhcb, in the following called \textit{default Kalman}, is significantly too slow. It relies on lookup tables for the magnetic field and the material distribution of the detector\cite{Bos:1070314}, so-called \textit{maps}. In addition it uses Runge-Kutta methods to solve the differential equations necessary to propagate the particle through the regions with an inhomogeneous magnetic field. Accessing the values in the lookup table and solving the differential equations are time consuming and prohibit the usage of the current Kalman filter in the first stage of the upgraded trigger system. This conclusion is independent of the choice of computing architecture (CPU or GPU) which is used for the first trigger stage.

In this paper, a fully parametrized version of the Kalman filter in \lhcb, called \textit{parametrized Kalman}, is presented. It obtains precise values of track parameters and track quality variables, 
while relying on neither computationally costly extrapolation methods nor material or magnetic field maps.

\section{Detector and simulation}
\label{sec:Detector}

The \lhcb detector~\cite{LHCb-DP-2008-001} is a single-arm forward
spectrometer covering the \mbox{pseudorapidity} range $2<\eta <5$.
Its Run 3 configuration includes a high-precision tracking system
consisting of a silicon-pixel vertex detector surrounding the $pp$
interaction region~\cite{LHCb-TDR-013} (VELO), a large-area silicon-strip detector (Upstream Tracker (UT))~\cite{LHCb-TDR-015} located
upstream of a dipole magnet with a bending power of about
$4{\mathrm{\,Tm}}$~\cite{LHCb-TDR-001}, and three stations of scintillating-fibre detectors (SciFi)~\cite{LHCb-TDR-015} 
placed downstream of the magnet. Different types of charged hadrons are distinguished using information
from two ring-imaging Cherenkov detectors~\cite{LHCb-TDR-003,LHCb-TDR-014}.
Photons, electrons and hadrons are identified by a calorimeter system consisting of an electromagnetic
and a hadronic calorimeter~\cite{LHCb-TDR-002,LHCb-TDR-014}. Muons are identified by a
system composed of alternating layers of iron and multiwire proportional chambers~\cite{LHCb-TDR-004,LHCb-TDR-014}.

Given the lack of collision data at this point for Run 3, simulation is required to model the effects of the detector response, the detector acceptance and the imposed selection requirements. In the simulation, $pp$ collisions are generated using \pythia~\cite{Sjostrand:2007gs,*Sjostrand:2006za} with a specific \lhcb configuration~\cite{LHCb-PROC-2010-056}.  Decays of unstable particles are described by \evtgen~\cite{Lange:2001uf}, in which final-state radiation is generated using \photos~\cite{Golonka:2005pn}. The interaction of the generated particles with the detector, and its response, are implemented using the \geant toolkit~\cite{Allison:2006ve, *Agostinelli:2002hh} as described in Ref.~\cite{LHCb-PROC-2011-006}.

\section{Principles}
\label{sec:Principles}
In the following, the Kalman filter formalism and its application in the \lhcb track reconstruction is outlined. During Kalman filtering, the information from measurements at detector planes is successively combined to obtain optimal estimates of the track parameters. 
The track is represented as a set of states at fixed $z$-positions\footnote{The detector coordinate system is chosen such that the $z$-axis is parallel to the beam line and charged particles are deflected in the direction of the $x$-axis.}, which are typically detector layers. Each of these states is given by $\boldsymbol{x}=(x,y,t_x,t_y,\frac{q}{p})$ and the corresponding covariance matrix $\boldsymbol{P}$, where $t_x$ and $t_y$ are the slopes with respect to the $z$ axis, $q$ the charge of the particle in units of the electron charge and \ptot its absolute momentum. 

The Kalman filter procedure needs an estimate of a state as a starting point. Filtering is then a repeated application of two steps. Firstly, the current state is extrapolated to the next detector layer, and secondly, the extrapolated state is updated using the measurement in this layer. If the track has no associated measurement in this layer, the update step is omitted. 
These steps can be formalized as follows: given the state {($\boldsymbol{x}_{k-1|k-1}$}, $\boldsymbol{P}_{k-1|k-1}$) at position $z_{k-1}$, the extrapolated state ($\boldsymbol{x}_{k|k-1}$, $\boldsymbol{P}_{k|k-1}$) at position $z_{k}$ is given by
\begin{align}
\boldsymbol{x}_{k|k-1} &= \boldsymbol{f}_k(\boldsymbol{x}_{k-1|k-1}),\\
\boldsymbol{P}_{k|k-1} &= \boldsymbol{F}_k\boldsymbol{P}_{k-1|k-1}\boldsymbol{F}_k^T+\boldsymbol{Q}_k,
\end{align}
where the extrapolation function $\boldsymbol{f}_k(\boldsymbol{x})$ is given by five individual mappings $\boldsymbol{f}_k=(f_k^x,f_k^y,f_k^{t_x},f_k^{t_y},f_k^{\frac{q}{p}})$. This leads to the transport matrix $\boldsymbol{F}_k$ as
\begin{align}
F_k^{ij} = \frac{\partial f_k^i}{\partial x_j}.\label{eq:transp_matrix}
\end{align}
The noise matrix $\boldsymbol{Q}_k$ accounts for uncertainties of the extrapolation, \eg due to scattering at the material of the detector layers or the material in between. 

The extrapolated state is then combined with the measurement $\boldsymbol{m}_{k}$ in the respective detector layer to obtain the new state estimate at the position $\boldsymbol{z}_{k}$, $\boldsymbol{x}_{k|k}$ and $\boldsymbol{P}_{k|k}$, using the following steps:
\begin{align}
\boldsymbol{r}_k &= \boldsymbol{m}_k-\boldsymbol{H}_k\boldsymbol{x}_{k|k-1}, \\
\boldsymbol{S}_k &= \boldsymbol{H}_k\boldsymbol{P}_{k|k-1}\boldsymbol{H}_k^T + \boldsymbol{R}_k, \\
\boldsymbol{K}_k &= \boldsymbol{P}_{k|k-1}\boldsymbol{H}_k^T\boldsymbol{S}_k^{-1},\\
\boldsymbol{x}_{k|k} &= \boldsymbol{x}_{k|k-1} + \boldsymbol{K}_k \boldsymbol{r}_k,\\
\boldsymbol{P}_{k|k} &= (\boldsymbol{1}-\boldsymbol{K}_k\boldsymbol{H}_k)\boldsymbol{P}_{k|k-1}.
\end{align}
Here $\boldsymbol{H}_k$ projects the estimated state vector to the measurement space in order to allow a calculation of the residual $\boldsymbol{r}_k$. The covariance matrix of this residual is given by $\boldsymbol{S}_k$ and is combined with the covariance matrix of the state to obtain the Kalman gain $\boldsymbol{K}_k$. The latter defines then how the estimated state is modified by the residual. The variance of the residual is given by $\boldsymbol{R}_k$.

Starting at the most upstream measurement, the measurements are successively added and the track parameters updated until the last detector layer is reached. 
The same procedure is repeated starting at the most downstream measurement and successively including more upstream measurements. This yields two sets of states at every measurement position, which can be combined to obtain the respective optimal state. 

The quality of a track can be estimated by its $\chi^2_{\text{track}}$ value. The value at each measurement is given by:
\begin{align}
\chi^{2}_{k} &= \chi^2_{k-1} + \boldsymbol{r}_{k}^{T} \boldsymbol{P}_{k|k}^{-1} \boldsymbol{r}_{k},
\end{align}
and $\chi^2_{\text{track}}$ is then simply $\chi^{2}_{k}$ after all measurements have been added using the combined, optimal states.

The optimal state estimates and the measurement information can also be used to remove measurements that show a large separation from the fitted trajectory by having a large contribution to the $\chi^2_{\text{track}}$ value. They are therefore likely to be wrongly associated to the respective track, and are so-called \textit{outliers}. Once an outlier is removed, all Kalman filter steps are performed again. This procedure can be repeated until the maximum allowed number of outliers are removed, or no more outliers are present.

The above formalism is also the basis of the Kalman filter that is currently used for track fitting in the \lhcb experiment.
The extrapolation functions $\boldsymbol{f}_k$ are based on maps of the magnetic field along the trajectory and numerical models for the extrapolations. Their complexities range up to a fifth-order Runge-Kutta method. The noise matrices $\boldsymbol{Q_k}$ are obtained by a dedicated model for the multiple scattering and a map of the material traversed by the particle.

In the parametrized Kalman filter presented in this paper, these two costly steps are replaced by simple parametrizations. The extrapolation functions $\boldsymbol{f}_k$ are given by analytic expressions that allow a fast evaluation and calculation of the derivatives in Equation \ref{eq:transp_matrix}.
The noise matrices $\boldsymbol{Q}_k$ depend on the momentum of the particle and are parametrized by a few parameters per extrapolation step.

An important difference with respect to the default Kalman filter is the treatment of energy loss due to the interaction with the detector material. While the multiple scattering is taken directly into account, the energy loss is not part of the extrapolation functions $\boldsymbol{f}_k$, \ie $f_k^{\frac{q}{p}}$ is the unity transformation. This shortcoming is compensated by choosing the momentum of the state vectors to represent the momentum at the moment of production of the particle. Thereby, the extrapolation functions also take this initial momentum as input and thus indirectly take into account all energy loss that happened on average up to the respective detector layer. The only caveat being that $\frac{q}{p}$ after the filtering is only the best representation of the true value at the production point of the particle.

\section{Parametrizations}
\label{sec:Parametrizations}
Depending on the strength of the magnetic field and the typical distance between detector layers, different empirical analytical functions for the extrapolation are used.

Inside the \velo, where the magnetic field is very weak, these functions and the noise matrix are given by:
\begin{align}
\renewcommand*{\arraystretch}{1.2}
\boldsymbol{f}(\boldsymbol{x}) =
\begin{pmatrix}f^x(\boldsymbol{x}) \\
  f^y(\boldsymbol{x}) \\
  f^{t_x}(\boldsymbol{x}) \\
  f^{t_y}(\boldsymbol{x}) \\
  f^{\frac{q}{p}}(\boldsymbol{x})
\end{pmatrix}
 = 
\begin{pmatrix}
  x+0.5[t_x+f^{t_x}(\boldsymbol{x})]\Delta z\ \\
  y+t_y\Delta z\\
  t_x+p^{\text{V}}_0\frac{q}{p}(z_0+p^{\text{V}}_1)\Delta z\\
  t_y\\
  \frac{q}{p}
\end{pmatrix}
\end{align}
and
\begin{align}
\boldsymbol{Q} =
\begin{pmatrix}
  \left(\tilde{p}^{\text{V}}_1\Delta z\right)^2Q^{t_xt_x} & 0 & \tilde{p}^{\text{V}}_2\sqrt{Q^{xx}Q^{t_xt_x}} & 0 & 0\\
  0 & \left(\tilde{p}^{\text{V}}_1\Delta z\right)^2Q^{t_yt_y} & 0 & \tilde{p}^{\text{V}}_3\sqrt{Q^{yy}Q^{t_yt_y}} & 0\\ 
  \tilde{p}^{\text{V}}_2\sqrt{Q^{xx}Q^{t_xt_x}} & 0 & \left(\tilde{p}^{\text{V}}_0\left|\frac{q}{p}\right|\right)^2 & 0 & 0\\
  0 & \tilde{p}^{\text{V}}_3\sqrt{Q^{yy}Q^{t_yt_y}} & 0 & \left(\tilde{p}^{\text{V}}_0\left|\frac{q}{p}\right|\right)^2 & 0\\
  0 & 0 & 0 & 0 & 0\\
\end{pmatrix},\label{eq:noisematrix}
\end{align}
where $\Delta z$ is the extrapolation distance along the $z$-direction and $z_0$ the initial or final $z$ coordinate for a downstream or upstream extrapolation, respectively. The parameters $p^{\text{V}}_0$, $p^{\text{V}}_1$ and $\tilde{p}^{\text{V}}_0$ to $\tilde{p}^{\text{V}}_3$ are the same for all upstream and downstream extrapolations inside the \velo.
They are determined using simulated \decay{\Bs}{\phi\phi} decays within the \lhcb software framework, where \decay{\phi}{\Kp\Km}. This simulated sample allows to create a dataset $D$, containing pairs of states representing two consecutive measurements of one track inside the \velo. In addition to the true state parameters obtained from the simulation, also an extrapolation of each state to the $z$ position of the respective other state is included in the dataset. Such extrapolation is based on the default extrapolation algorithm in \lhcb~\cite{Bos:1070314}.
This dataset allows tuning the parameters employing a minimization of the following likelihood-inspired function:
\begin{align}
\prod_{D}\left[\mathcal{G}\left(f^s(\boldsymbol{x_1})-\boldsymbol{x_2}^s,\sqrt{Q^{ss}}\right)+ c\right].
\end{align}
Here, $\mathcal{G}(x,\sigma_x)$ is a normalized Gaussian distribution centered around $0$ with width $\sigma_x$. The two states of each dataset entry are represented by $\boldsymbol{x}_1$ and $\boldsymbol{x}_2$, and the variable $s$ is one of the state variables, $s\in\{x,t_x,y,t_y\}$. The positive empirical constant $c$ is chosen to be small with respect to the amplitude of the Gaussian function and softens the impact of outliers.

In a first step, the extrapolation functions $f^x$ to $f^{t_x}$ are tuned individually, taking into account that $f^x$ depends on the previously determined parameters for $f^{t_x}$. These tuning minimizations employ the state vector $\boldsymbol{x}_2$ that is obtained by the extrapolation of the state vector $\boldsymbol{x}_1$. This choice improves the precision of the parametrized extrapolation, by removing the effect of multiple scattering that would be present if instead the true state was chosen for $\boldsymbol{x}_2$.

In a second step, the parameters of the extrapolation functions are fixed, and a minimization of the following function is performed:
\begin{align}
\prod_{D}\left[\mathcal{G}_2\left(
f^d(\boldsymbol{x_1})-\boldsymbol{x_2}^d,
f^{t_d}(\boldsymbol{x_1})-\boldsymbol{x_2}^{t_d},
\sqrt{Q^{dd}},
\sqrt{Q^{t_dt_d}},
Q^{dt_d}/\sqrt{Q^{dd}Q^{t_dt_d}}\right)+ c\right].
\end{align}
Here, $\mathcal{G}_2(x,y,\sigma_y,\sigma_y,\rho)$ is a normalized two-dimensional Gaussian distribution centered around $0$ with widths $\sigma_x$ and $\sigma_y$ and a correlation factor $\rho$. The variable $d$ is either $x$ or $y$. In this minimization, the true state vector $\boldsymbol{x}_2$ is used in order to get the correct estimate of the parameters for the respective elements of the noise matrix $\boldsymbol{Q}$.

Inside the UT and the SciFi detector stations, the magnetic field is significantly stronger than inside the \velo and higher order terms are needed for the extrapolation functions:
\begin{align}
\boldsymbol{f}(\boldsymbol{x}) = 
 \renewcommand*{\arraystretch}{1.4}
 \begin{pmatrix}
 x+\left[p^{\text{T}}_3t_x+(1-p^{\text{T}}_3)f^{t_x}(\boldsymbol{x})\right]\Delta z\\
 y+\left[p^{\text{T}}_5t_y + (1-p^{\text{T}}_5)f^{t_y}(\boldsymbol{x})\right]\Delta z\\
 t_x+\left[p^{\text{T}}_0\frac{q}{p} + p^{\text{T}}_1(\frac{q}{p})^3 + p^{\text{T}}_2y^2\frac{q}{p}\right]\Delta z\\
 t_y + p^{\text{T}}_4\frac{q}{p}t_x\frac{y}{|y|}\\
 \frac{q}{p}
 \end{pmatrix}.
\end{align}
The noise matrix is given in full analogy to Equation \ref{eq:noisematrix} with the parameters $\tilde{p}^{\text{T}}_0$ to $\tilde{p}^{\text{T}}_3$, where T either stands for the UT or the SciFi detector. These parameters and the parameters $p^{\text{T}}_0$ to $p^{\text{T}}_4$ are individually determined on simulation for every step from one detector layer to the next and for the upstream and downstream extrapolation separately. The same strategy as for the tuning of the parameters related to the extrapolation inside the \velo is followed.

For the long extrapolations between the different tracking subdetectors, more sophisticated parametrizations are necessary. In the case of the step between the \velo and the UT, where the magnetic field is still weak, the extrapolation is based on two equations. The first describes the change in momentum along the $x$-direction of the particle:
\begin{align}
\Delta p_x 
= p\left(\frac{t_{x,\text{UT}}}{\sqrt{1+t^2_{x,\text{UT}}+t^2_{y,\text{UT}}}}-\frac{t_{x,\text{V}}}{\sqrt{1+t^2_{x,\text{V}}+t^2_{y,\text{V}}}}\right)
= q\int \left( \text{d}\boldsymbol{l}\times \boldsymbol{B} \right)_{x},\label{eq:delta_px}
\end{align}
where $t_{x/y,\text{UT}}$ and $t_{x/y,\text{V}}$ are the state variables at the first UT detector layer and the last measurement inside the \velo, respectively. The right hand side of the equation consists of an integral of the magnetic field along the trajectory of the particle. Note that the integral expression is simply a parameter which was fitted for on the dataset.
The second ingredient for the extrapolation is to model the effect of the magnetic field as a single kink of the trajectory at a certain $z$-position $z_{\text{mag}}$ between the \velo and the UT:
\begin{align}
x_{\text{UT}} = x_{\text{V}} + (z_{\text{mag}}-z_{\text{V}})t_{x,\text{V}} + (z_{\text{UT}}-z_{\text{mag}})t_{x,\text{UT}},\label{eq:kink}
\end{align}
where $z_{\text{V}}$ and $z_{\text{UT}}$ are the positions of the states inside the \velo and the UT, respectively.

Equation \ref{eq:delta_px} can be solved for $t_{x,\text{UT}}$ and Equation \ref{eq:kink} is then employed to get an expression for $x_{\text{UT}}$. The unknowns in these expressions are parametrized as a function of the state variables inside the \velo:
\begin{align}
t_{y,\text{UT}} &= t_{y,\text{V}} + p^{\text{S}}_0\frac{q}{p}t_{x,\text{V}}\frac{y_{\text{V}}}{|y_{\text{V}}|}\\
\int \left( \text{d}\boldsymbol{l}\times \boldsymbol{B} \right)_{x} &= p^{\text{S}}_1 + p^{\text{S}}_2z_{\text{V}} + p^{\text{S}}_3 t^2_{y,\text{V}} \\
z_{\text{mag}} &= p^{\text{S}}_4 + p^{\text{S}}_5 z_{\text{V}} + p^{\text{S}}_6z^2_{\text{V}} + p^{\text{S}}_7 t^2_{y,\text{V}}.
\end{align}
In addition, the $y$-position of the extrapolated state is given by:
\begin{align}
y_{\text{UT}} = y_{\text{V}}+\left[p^{\text{S}}_8t_{y,\text{V}} + (1-p^{\text{S}}_8)t_{y,\text{UT}}\right]\Delta z,
\end{align}
where $\Delta z$ is defined as the difference between $z_{\text{UT}}$ and $z_{\text{V}}$. The noise matrix is defined in analogy to Equation \ref{eq:noisematrix} with the parameters $\tilde{p}^{\text{S}}_0$ to $\tilde{p}^{\text{S}}_3$. These parameters and the parameters $p^{\text{S}}_0$ to $p^{\text{S}}_8$ are individually determined for the upstream and downstream extrapolation.
The same strategy as for the tuning of the parameters related to the extrapolation inside the \velo is followed.

The extrapolation from the UT to the SciFi detector is more delicate
because it is done over a distance of more than 5 meters through
a strong magnetic field. Moreover, this field is far from uniform -
in particular, it varies rapidly in the upper and lower regions,
close to the magnet yoke. To ensure a good quality of the global
track fit, the error on the extrapolation should be well
below the other sources of error, mainly multiple
scattering. The chosen solution is an expansion of the magnetic
deviation in powers of $q/p$. The parametrization aims at giving good precision for charged particles
used in physics analyses, that is for trajectories which roughly come from
the origin. 

To do so, the \textit{ideal} direction $(t_x^0,t_y^0)$ as the one of a
particle of charge $q$, momentum $p$, starting from the origin and
hitting the UT detector layer in a given point $(x,y)$ is defined. As a good approximation, we can take $t_x^0 = x/z+{\cal B}q/p$, $t_y^0 = y/z$, where $\cal{B}$ is proportional to the integrated field between the origin and the UT. The deviations from the ideal direction, $\delta t_x=t_x-t_x^0$, $\delta t_y=t_y-t_y^0$, are small, so only a first order expansion in  $\delta t_x,\delta t_y$ is considered. Corrections of higher order would be negligible compared to multiple scattering errors.

Finally, a polynomial expansion in $q/p$ for the ideal direction is built, and a correction
in $\delta t_x,\delta t_y$ with coefficients which are themselves polynomials of $q/p$ is added:
\begin{align}
 f^x(\boldsymbol{x}) = x+t_x\Delta z+ \sum_{k=1}^{K_1} A^x_k(x,y)\left(\frac{q}{p}\right)^k
 +\sum_{k=1}^{K_2} \left(B^x_k(x,y)\,\delta t_x+C^x_k(x,y)\,\delta t_y\right)\left(\frac{q}{p}\right)^k,
 \end{align}
where the first two terms are the straight line extrapolation, and the
next ones the curvature correction. Similar expressions are used for
the other state parameters $f^y(\boldsymbol{x})$,
$f^{t_x}(\boldsymbol{x})$, $f^{t_y}(\boldsymbol{x})$. The
degrees of expansion $K_1$ and $K_2$ are tuned for each parameter to
obtain the required precision. In practice $K_1=9$, $K_2=7$
for $f^x$ and $f ^{t_x}$ and  $K_1=7$, $K_2=5$
for $f^y$ and $f ^{t_y}$ are used.

The dependence on $x,y$ of the coefficients $A^u_k$, $B^u_k$, 
$C^u_k$, with $u=x,y,t_x,t_y$, is described through a tabulation on a grid of 50$\times$50
points regularly spaced on the rectangle defined by $|x/z|\leq 0.25$,
$|y/z|\leq 0.25$, by steps $\Delta X$, $\Delta Y$. In order to avoid a systematic convexity bias of a
bilinear interpolation, the values at $x,y$ are computed by a quadratic
interpolation between the tabulated values at the six closest points on
the grid: if $(X,Y)$ is the closest one, these values are: $F_{00}=(X,Y)$, $F_{+0}=F(X+\Delta X,Y)$, $F_{-0}=F(X-\Delta X,Y)$,
$F_{0+}=F(X,Y+\Delta Y)$, $F_{0-}=F(X,Y-\Delta Y)$, 
and $F_{\varepsilon_x \varepsilon_y}=F(X+\varepsilon_x\Delta X,Y+\varepsilon_y\Delta Y)$,
 where $\varepsilon_x$ and $\varepsilon_y$ are the signs of $\xi=(x-X)/\Delta X$
and $\psi=(y-Y)/\Delta Y$, respectively. With these notations the
interpolation formula for a quantity $F$ is given by :
\begin{align}
F(x,y) = F_{00} +F_d\,\xi\psi + \big(&(F_{+0}-F_{-0})\,\xi +  (F_{0+}-F_{0-})\,\psi \nonumber \\
 &+ (F_{+0}+F_{-0}-2F_{00})\,\xi^2 +(F_{0+}+F_{0-}-2F_{00})\,\psi^2\big)/2\\
\text{with}\;\;\; F_d = &\varepsilon_x\varepsilon_y (F_{00}+F_{\varepsilon_x\varepsilon_y}-F_{\varepsilon_x 0}-F_{0 \varepsilon_y} ).
\end{align}
The tabulated values are obtained using the standard Runge-Kutta method of order 4, 
with 20
values of $q/p$ in the range ($-1/p_{min},1/p_{min}$), with
$p_{min}=3000 \mevc$ and a polynomial fit in $q/p$. As a consequence,
they do not give a reliable result for momenta below
$p_{min}$. Another limitation is the larger errors on the edges of the
acceptance, especially for $|t_y|\simeq 0.25$, where the field has
strong spatial variations.

\section{Performance}
\label{sec:Performance}
A sample of simulated proton-proton collisions that include a $B^0_s\rightarrow\phi\phi$, \decay{\phi}{\Kp\Km} decay
is used to compare the reconstruction quality of the parametrized and the default Kalman filter. The extrapolation of the most upstream state estimate to the beam line is the same in both filters and is based on a simplified material map of the detector \cite{Bos:1070314}. Therefore, not the state near the beam line, but the state at the most upstream measurement is employed for the comparison of the two Kalman filters. Although only tracks with measurements in each of the subdetectors are considered for this study, 
this is in principle not a requirement for operating the parameterized Kalman filter

\begin{figure}
\begin{center}
\includegraphics[width=.49\textwidth]{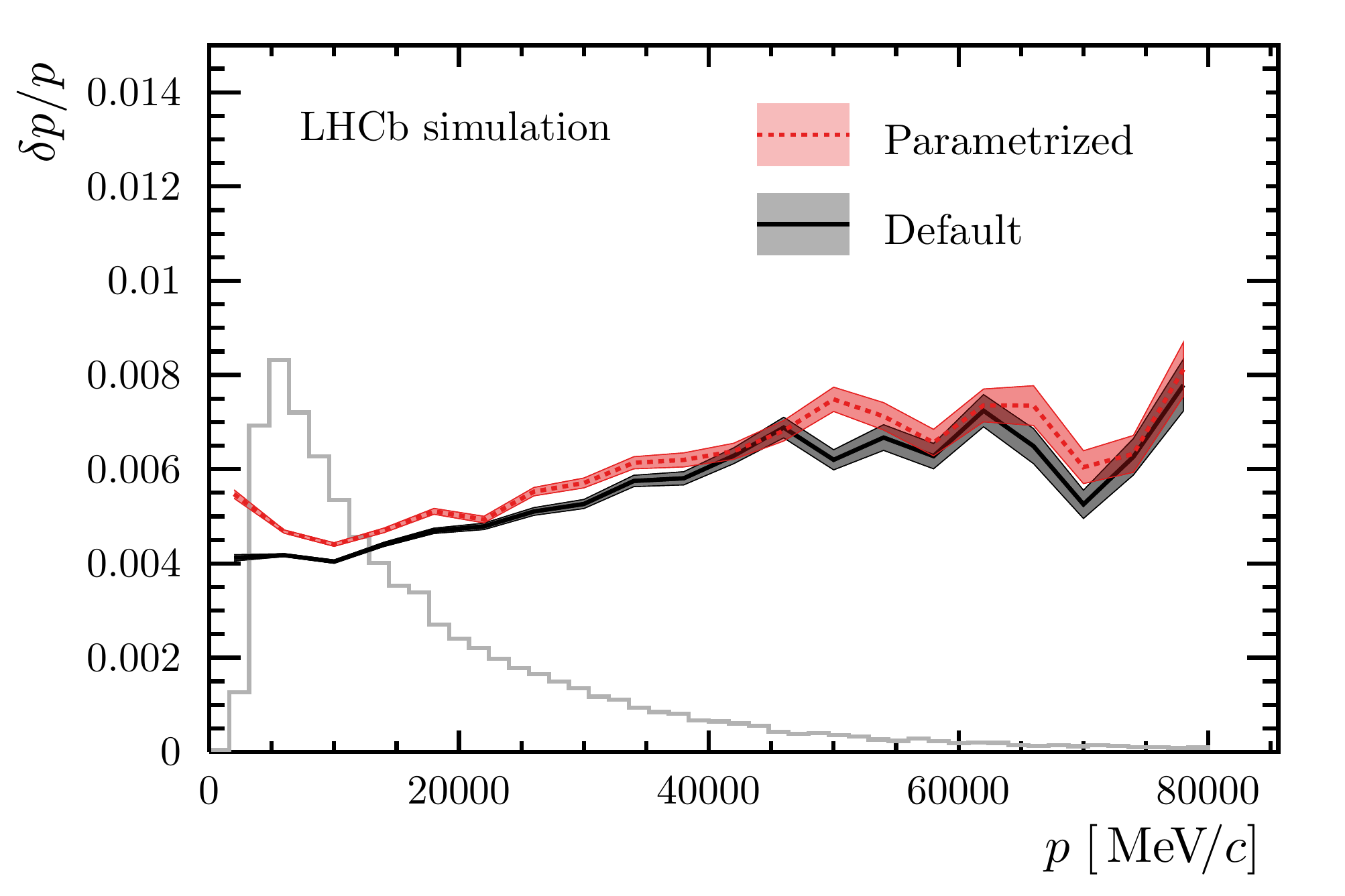}
\includegraphics[width=.49\textwidth]{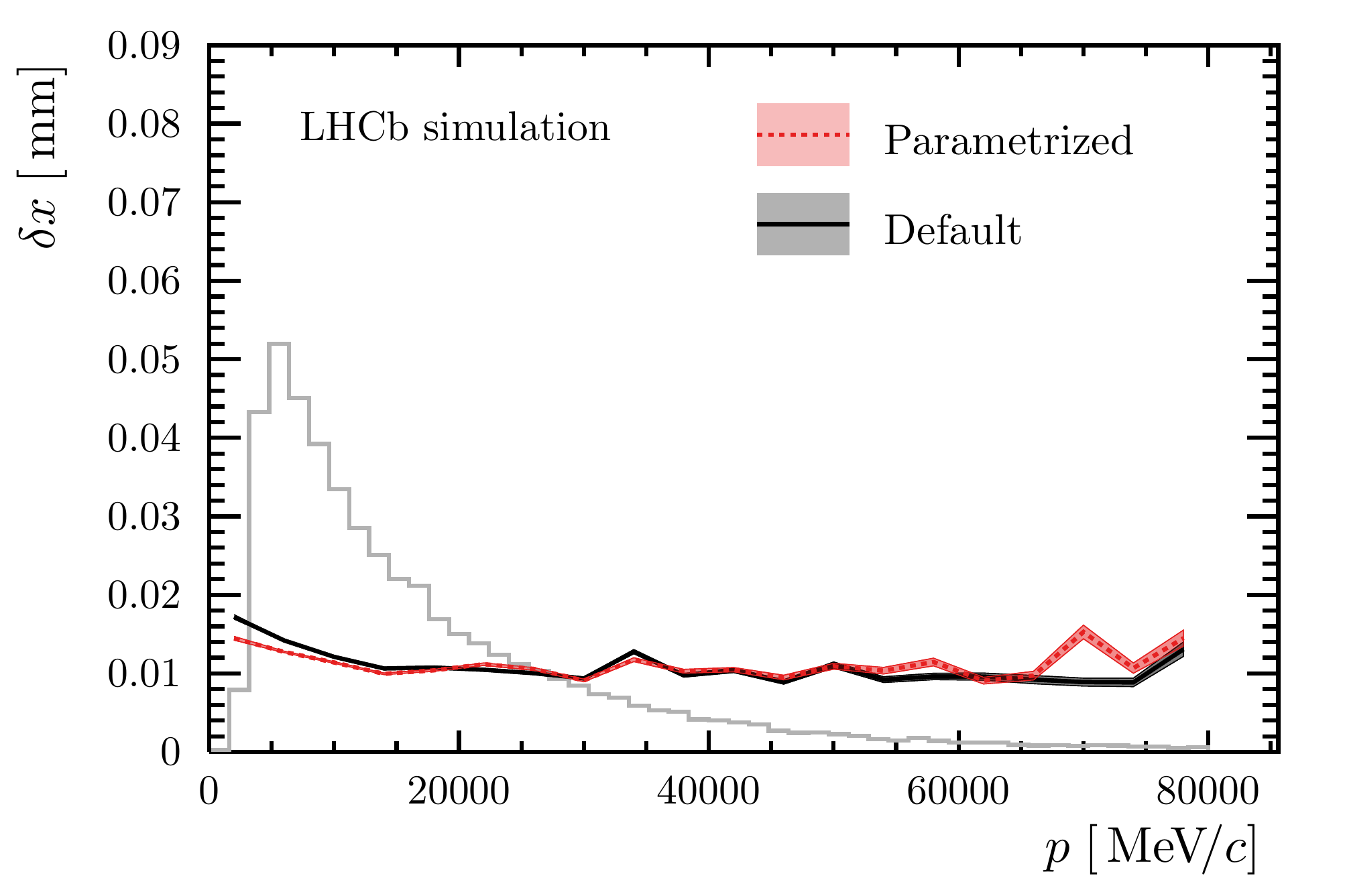}\\
\includegraphics[width=.49\textwidth]{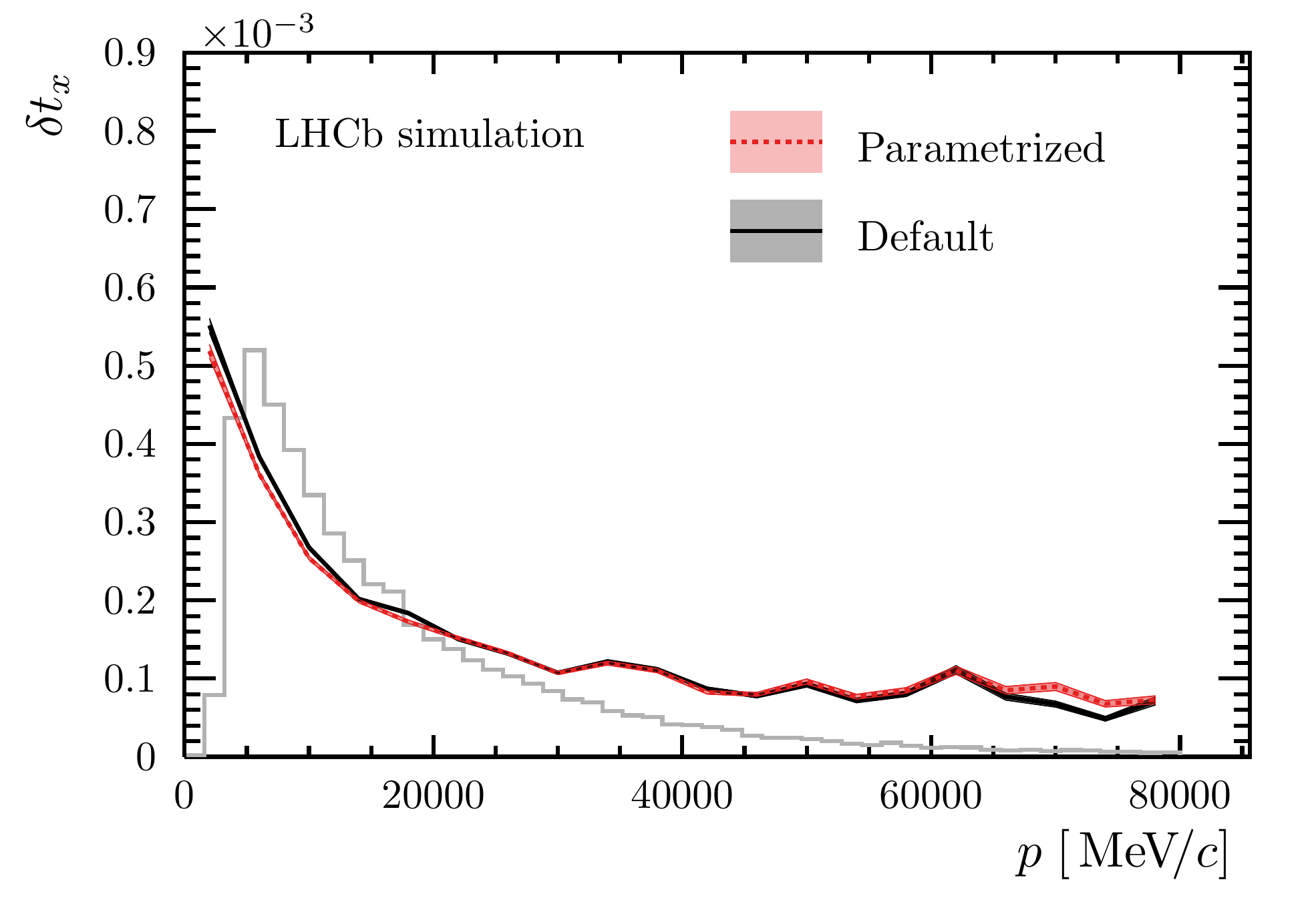}
\end{center}
\caption{\label{fig:Resolution}Comparison of the resolution in simulation in (top left) momentum, (top right) $x$-position and (bottom) slope $t_x$ between the default and parametrized Kalman filter. The resolution is represented by the root mean square of the residual distribution when comparing to the true value.}
\end{figure}

Figure \ref{fig:Resolution} compares the resolution of the momentum, the $x$-position and the slope $t_x$ as a function of the true momentum of a particle. 
Since the position and slope are nearly exclusively determined by the measurements in the VELO, where only a very weak magnetic field is present, the parametrizations of the parametrized Kalman filter are sufficient to obtain
results comparable to the default Kalman filter in these variables. In contrast, the momentum estimate strongly depends on the extrapolations in regions with strong magnetic field. There, especially at momenta below 10\gevc, an up to $20\%$ worse resolution is observed for the parametrized Kalman filter.

\begin{figure}
\begin{center}
\includegraphics[width=.49\textwidth]{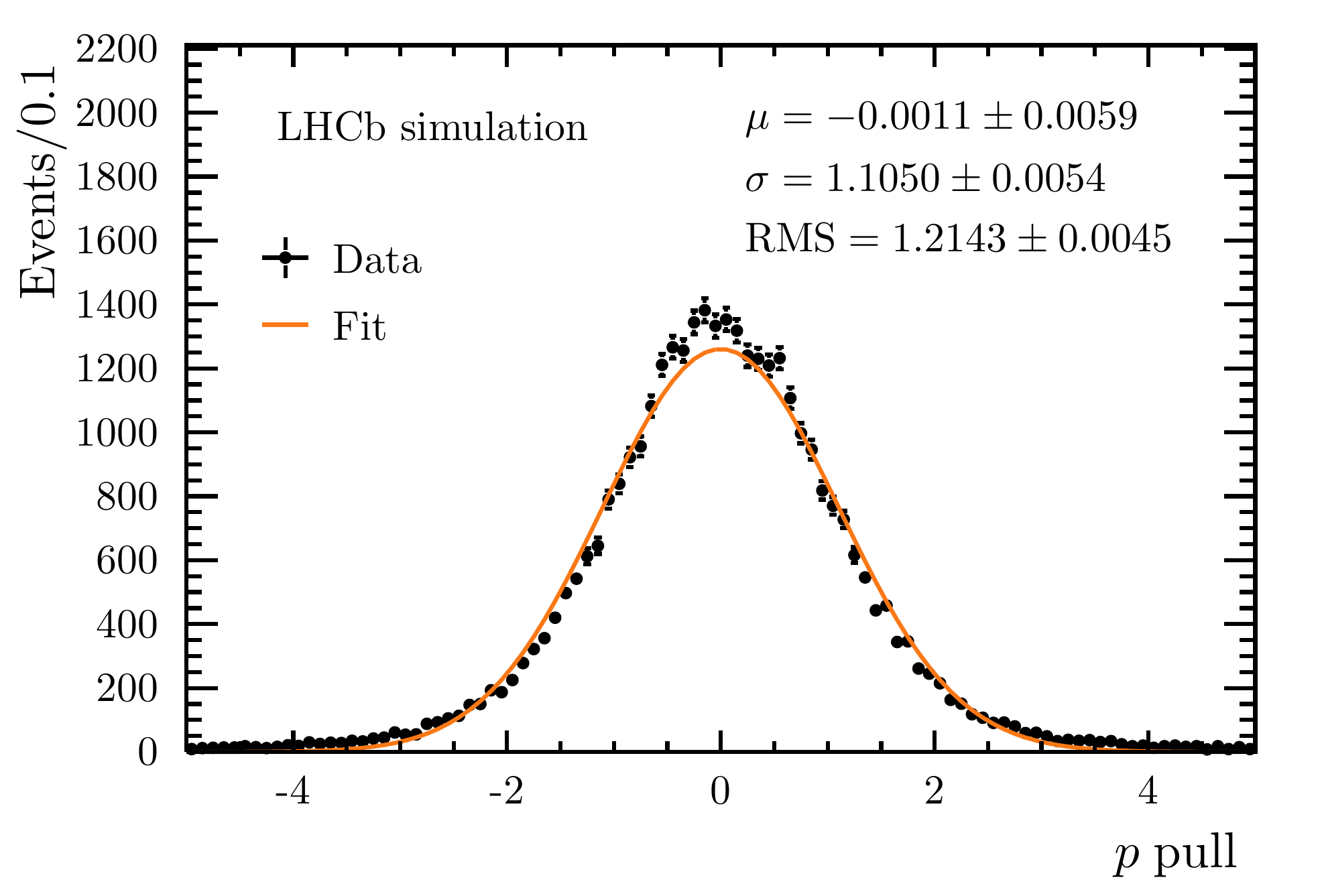}
\includegraphics[width=.49\textwidth]{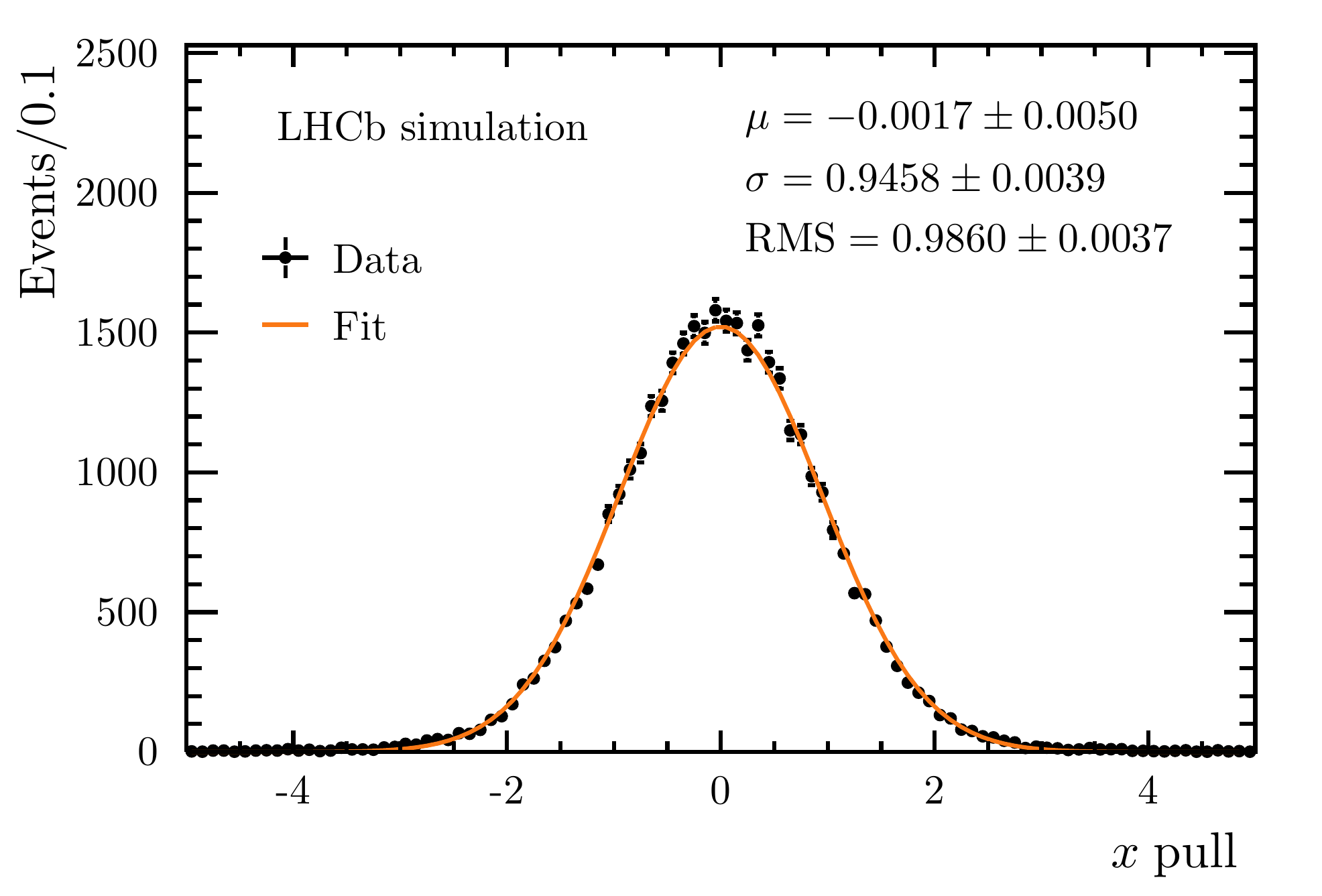}\\
\includegraphics[width=.49\textwidth]{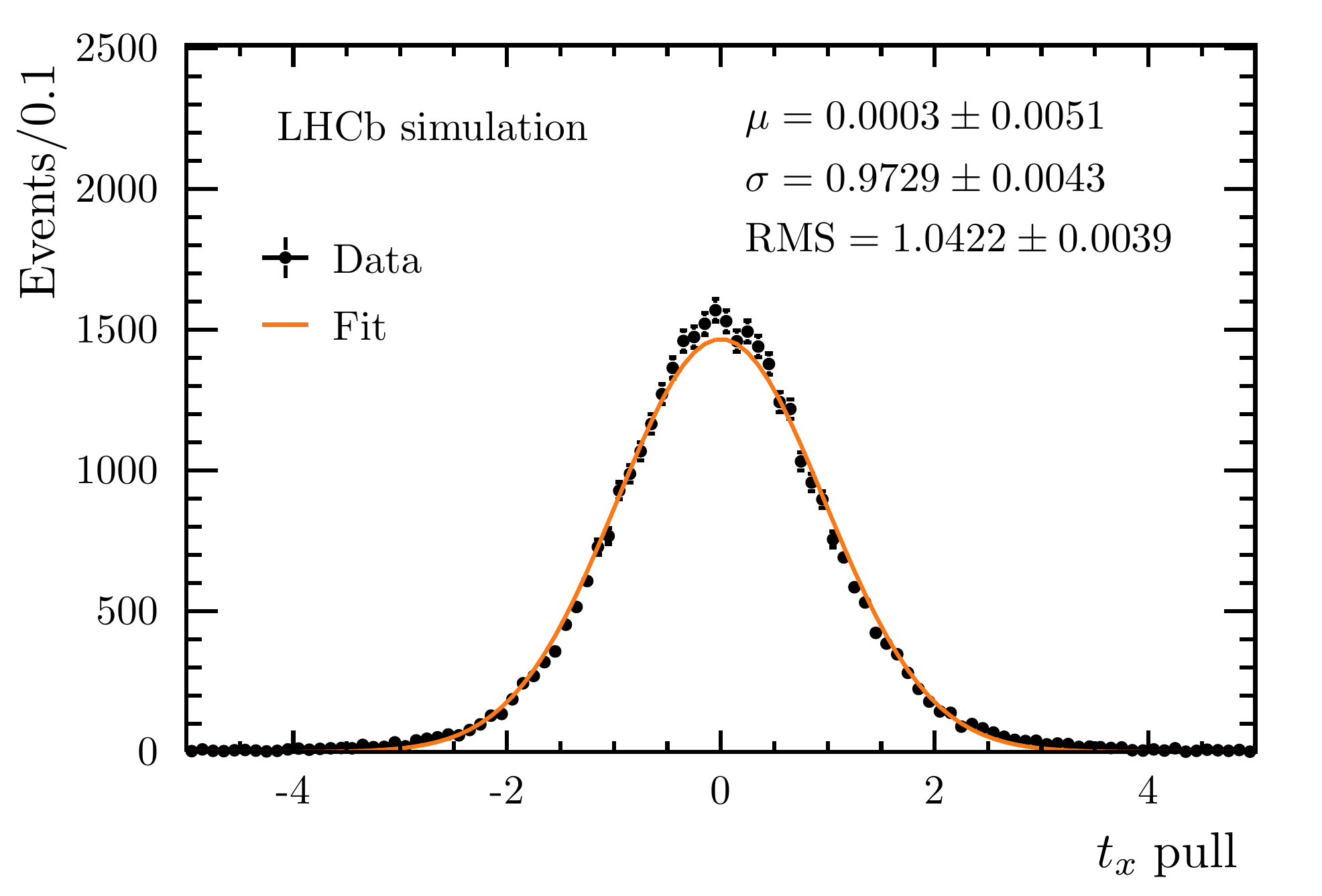}
\end{center}
\caption{\label{fig:Pull}Pull distributions of the momentum, $x$-position and slope $t_x$ estimates of the parametrized Kalman filter at the most upstream measurement. The given values correspond to the mean, width and root mean square of a Gaussian function that is fitted to the distribution.}
\end{figure}

The Kalman filter does not only provide an estimate of the state parameters, but also a corresponding covariance matrix. In Figure \ref{fig:Pull} the pull distributions of the estimated momentum, $x$-position and slope $t_x$ for the parametrized Kalman filter are shown.
In all three cases, good uncertainty estimates are visible. However, in analogy to the observations made for the resolution, the pull distribution of the momentum features slightly more pronounced tails.

\begin{figure}
\includegraphics[width=0.5\textwidth]{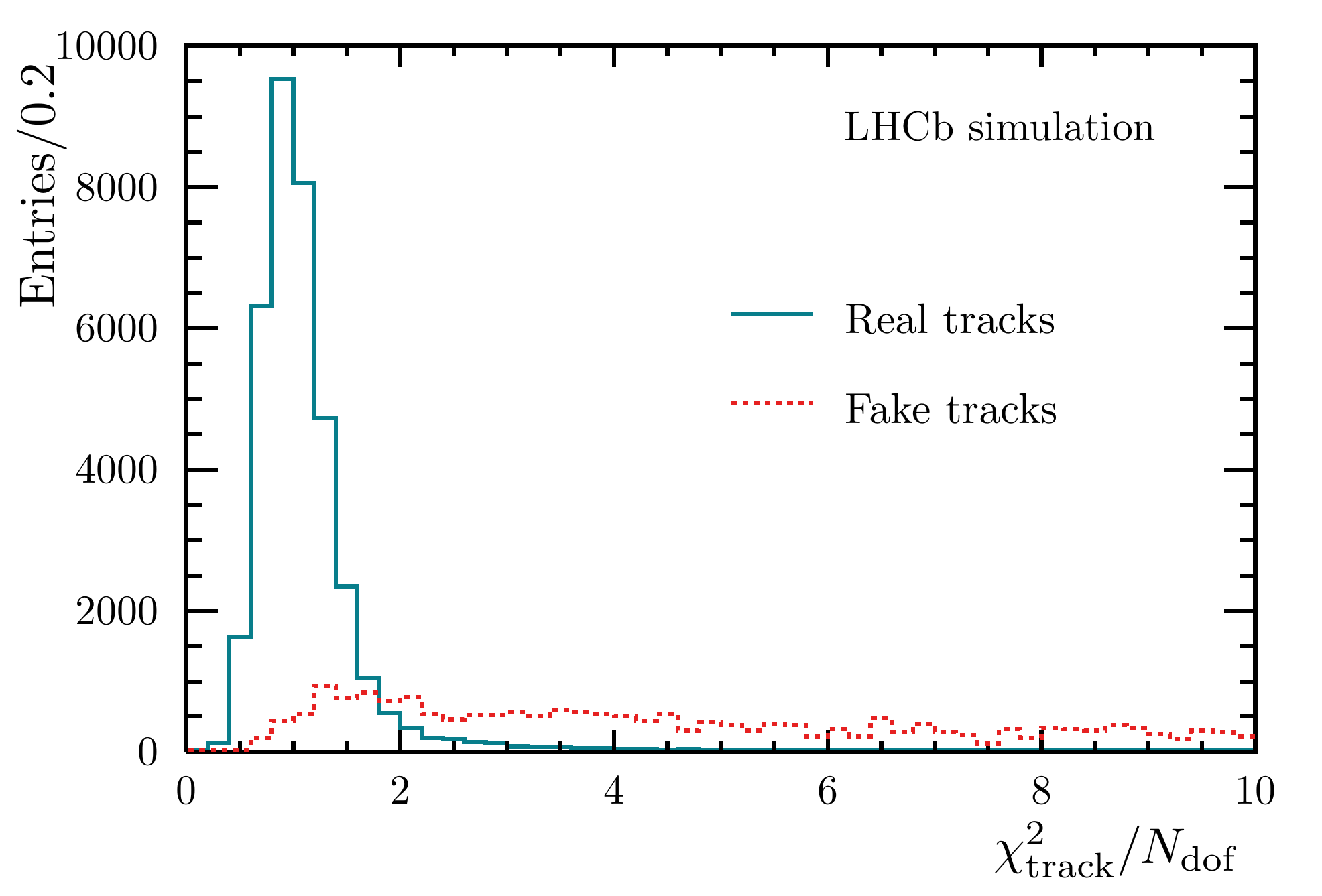}
\includegraphics[width=0.5\textwidth]{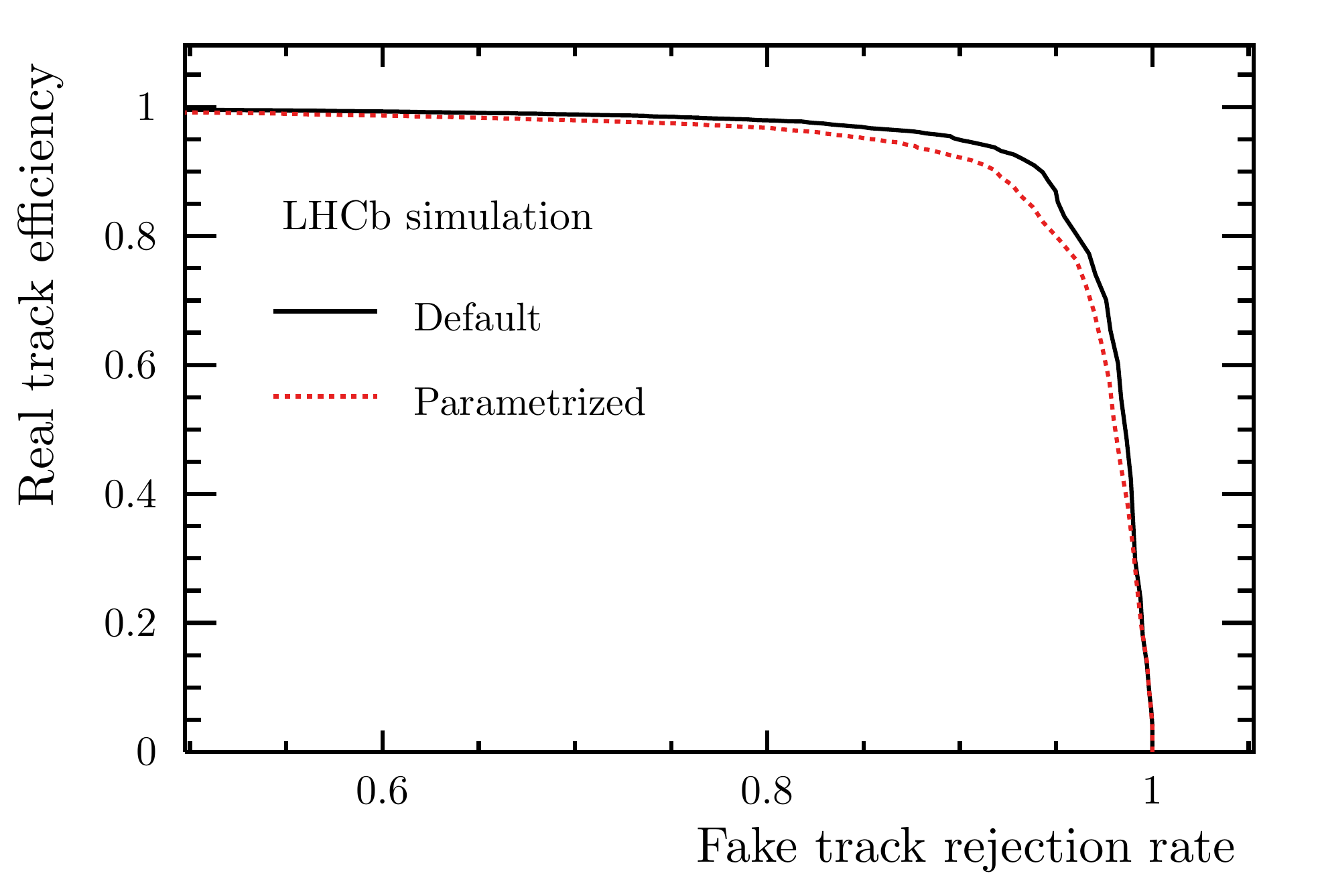}
\caption{\label{fig:Chi2}Track quality estimate, $\chi^2_{\text{track}}/N_{\text{dof}}$, in simulation for the parametrized filter (left). Fake tracks are shown in red and real tracks in black. Real track efficiency and fake track rejection for the parametrized and default Kalman filter (right).}
\end{figure}

Besides the estimate of the state near the beam line, which is used for the reconstruction of charged particles, an important output of the Kalman filter is the fit quality described by the $\chi^2_{\text{track}}$ per degrees of freedom $N_{\text{dof}}$. In Figure \ref{fig:Chi2}, this quantity is shown for the parametrized Kalman filter for real tracks coming from a particle and fake tracks consisting of random combinations of clusters. In addition, the real track efficiencies and fake track rejection rates are shown for both Kalman filter versions when applying upper bounds on this quantity.
The parametrized Kalman filter shows a slightly worse but overall comparable performance in separating the two track classes.

\begin{figure}
\includegraphics[width=0.5\textwidth]{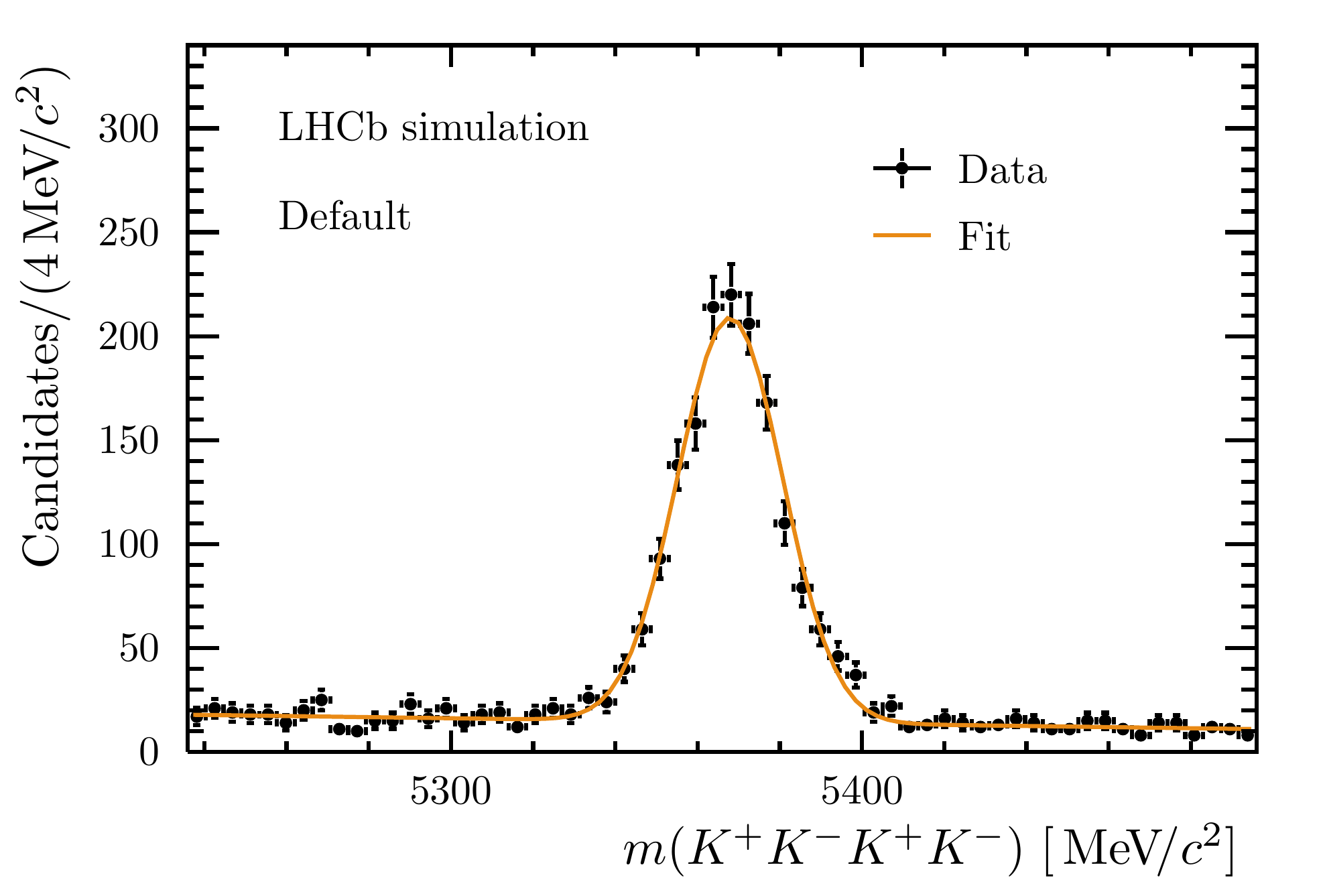}
\includegraphics[width=0.5\textwidth]{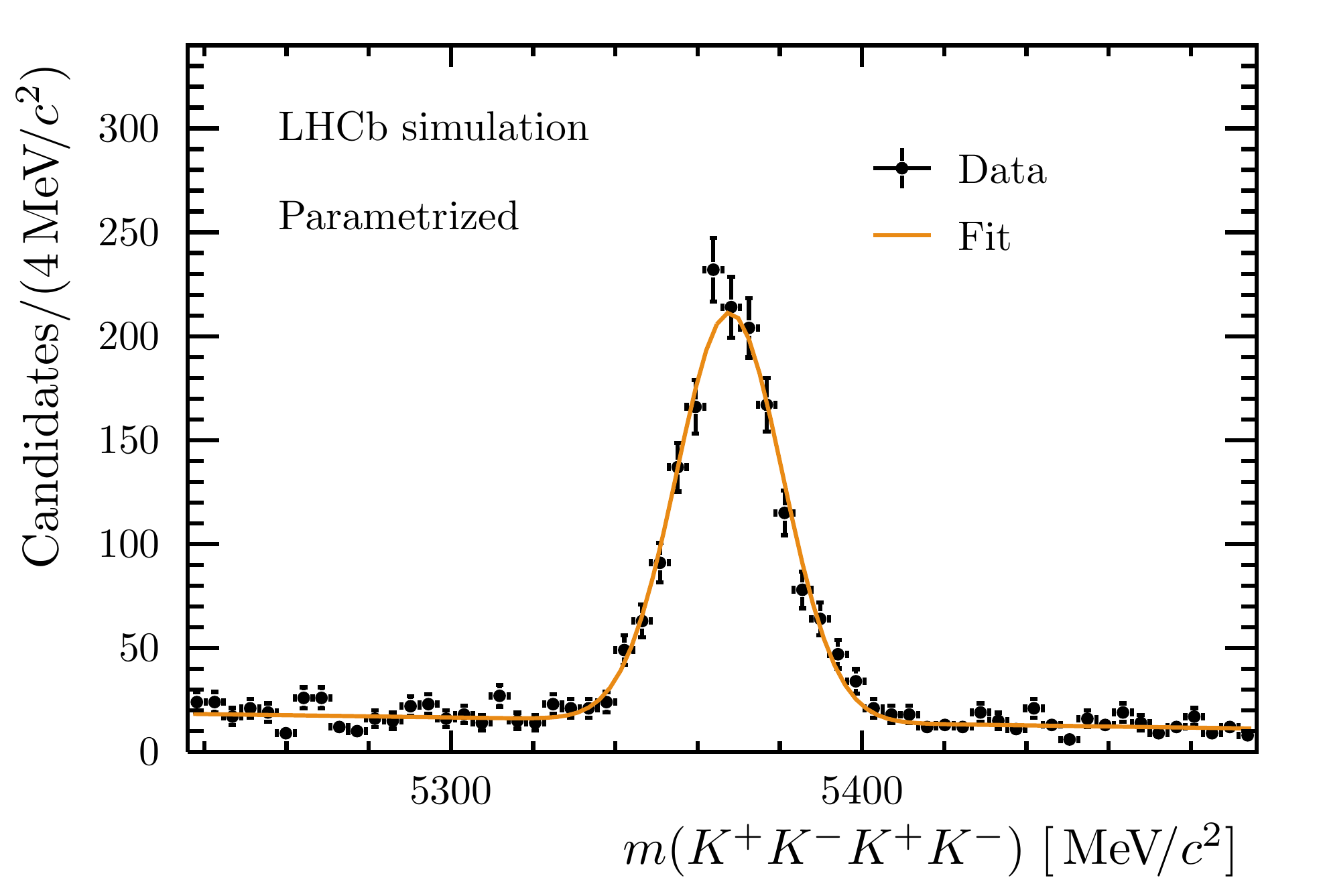}
\caption{\label{fig:mass}Reconstructed $B^0_s$ mass in simulated $B^0_s\rightarrow\phi\phi$ decays for the parametrized and the default Kalman filter. Fit projections are overlaid.}
\end{figure}

The fitted tracks are combined to reconstruct $B^0_s\rightarrow\phi\phi$ candidates. Figure \ref{fig:mass} shows the invariant mass distribution of candidates based on the two Kalman filter versions.
A single Gaussian distribution and a first order polynomial are employed to model the signal peak and the combinatorial background, respectively.
This yields nearly identical estimated mass resolutions of $12.8\mevcc$ and $12.9\mevcc$ for the default and the parametrized Kalman filter, respectively.

In order to compare the timing performance of the parametrized Kalman filter and the default Kalman filter, throughput studies on a machine with two Intel(R) Xeon(R) Silver 4214 processors were performed. Simulated proton-proton collisions 
were used in order to mimic the situation of real data taking. Depending on the configuration of the outlier removal strategy, an overall speedup factor between 4 and 5.5 with respect to the default Kalman filter was achieved. The largest speedup is achieved when no iterations for the outlier removal are performed. Singling out the calculation steps of the Kalman filter, \ie neglecting the part of the algorithms where the measurement information is constructed, the speedup factor is even larger and ranges from 5.7 to 10.

In the case of the parametrized Kalman filter, and singling out again the calculation step of the Kalman filter, $50\%$ of the time is spent extrapolating the states between the detector layers. Here, the extrapolation between the UT and the SciFi constitutes the biggest component with a relative fraction of $40\%$.
The remaining Kalman filter steps, consisting of updating the states with the cluster information and the combination of upstream and downstream filtered states, are responsible for  $16\%$ and $14\%$ of the time spent, respectively. The extrapolation to the beam line, which is based on the default LHCb extrapolation algorithm, is responsible for the remaining $20\%$ of the time budget.
\section{Conclusion}
\label{sec:Conclusion}
We presented an alternative implementation of a Kalman filter for the LHCb experiment. Based on simple parametrizations of material effects and the extrapolation through the magnetic field of the detector, this algorithm achieves a significant speedup with respect to the current implementation, while retaining comparable quality of the track parameters. In the future, further improvements of the parametrizations might allow an even better estimate of the track parameters and a subsequent speedup. Ideas currently under discussion include for example an analytic parametrization of the $x$ and $y$ dependence of the parameters employed in the extrapolation from the UT to the SciFi detector and a better account for the limited acceptance of low momentum particles.
The version presented in this document or a future implementation might therefore be well suited for the usage in the \lhcb software trigger system for Run 3 of the \lhc.
\section*{Acknowledgements}
The authors would like to thank the LHCb computing and simulation teams for their support and for producing the simulated LHCb samples used in the paper. We also would like to thank the LHCb RTA team for supporting this publication and reviewing the work. 
M. De Cian acknowledges support from the Swiss National Science Foundation grant ``Probing right-handed currents in quark flavour physics", PZ00P2\_174016.


\addcontentsline{toc}{section}{References}
\bibliographystyle{LHCb}
\bibliography{main,standard,LHCb-PAPER,LHCb-CONF,LHCb-DP,LHCb-TDR}


%
%
%
%

\end{document}